\documentstyle{article}

\input{psfig}

\begin{document}

\title{Baryon moments in a QCD-based model\footnote{\ \ Presented by L. Durand
at the Como Conference on {\em Quark Confinement and the Hadron Spectrum II}, 
June, 1996}}

\author{Loyal Durand and Phuoc Ha\\
Department of Physics, University of Wisconsin\\
Madison, Wisconsin 53706, USA}
\date{}
\maketitle

{\small\noindent We have analyzed the theory of the baryon
magnetic moments in the (approximate) QCD setting
suggested by Brambilla {\em et al.} By modifying their derivation of 
the $qqq$ interaction and wave equation for baryons, 
we derived expressions for the baryon moments in
terms of the underlying quark moments, including the first corrections
associated with the binding of the quarks in baryons. The results, which hold
in the ``quenched approximation'' in which the contributions
of virtual quark pairs are neglected, fail to describe the measured moments,
with typical errors on the order of 7\%. 
We conclude from our analysis
that the quenched approximation is at fault, and that the baryon moments
give a sensitive test of that standard approximation in the lattice and
analytic approaches to QCD.}

\section{Introduction}
The magnetic moments of the stable baryons are very well measured, and
their theoretical calculation gives a sensitive test of our understanding
of baryon structure in quantum chromodynamics (QCD). Although
the basic pattern and approximate magnitudes of the 
moments can be explained using the nonrelativistic
quark model, the deviation of the moments from the quark-model pattern
has not been explained dynamically despite many attempts. We discuss that 
problem here from the point of view of nonperturbative QCD. In particular,
we have analyzed the theory of the moments in the approximate QCD setting
developed in the work of Brambilla {\em et al.} \cite{brambilla}, who used 
a Wilson-loop approach to study the $qqq$ bound state
problem. By modifying their derivation of the $qqq$ interaction and wave
equation, we have derived expressions for the baryon moments in
terms of the underlying quark moments, including the first corrections
associated with the binding of the quarks in baryons. Our results hold in
the ``quenched'' approximation in which there are no internal quark loops
embedded in the world sheet swept out by the Wilson lines joining the
valence quarks, and no pairs associated with the valence lines.
The corrections to the moments are of relative order
$\langle V\rangle/m_q$, where $V$ is a typical component of the binding
potential and $m_q$ is a constituent quark mass, and
are potentially large enough to explain the deviations of the measured
moments from the quark-model values. 

To test the theory, we 
have constructed variational wave functions for the baryons using the 
interactions derived by Brambilla {\em et al.} \cite{brambilla}, including
all spin and orbital configurations possible for $J^P=\frac{1}{2}^+$ and 
internal and total orbital angular momenta $L\leq 2$, and used them to calculate the moments. The effects of excited orbital
contributions to the moments are negligible, as expected. 
The new contributions tend to
cancel, and the remnants do not have the correct pattern 
to explain the discrepancies
between theory and experiment. 

The approximations underlying the
Wilson-line analysis are known.  Since of these, only the quenched 
approximation has a direct effect on the spin dependent
terms with which we are dealing, we
conclude that the use of the quenched approximation is 
responsible for the deviations of theory from experiment, and
that the moment problem 
provides a sensitive test of this standard approximation in lattice
and analytic QCD.

In a world-sheet picture, the inclusion of internal quark loops which describe meson emission and absorption by the baryon would allow 
the introduction of internal orbital angular momentum and spin, and would 
affect the moments. We are now investigating meson-loop
contributions using approximations suggested by the world-sheet picture 
and chiral pertubation theory. In the following sections, we sketch our
derivations and calculations, and justify our conclusions. A more detailed
discussion will be given elsewhere.

\section{Baryon moments in QCD} 

\subsection{The problem}
The simple, nonrelativistic quark model gives a qualitatively good
description of the baryon moments. Under the assumption that each baryon
is composed of three valence or constituent quarks in a state with all
internal orbital angular momenta equal to zero, the moments are given
by expectation values of the spin moment operators
\begin{equation}
\mbox{\boldmath$\mu$}_{\rm QM}=\sum_{\rm q}\,\mu_{\rm q}
\mbox{\boldmath$\sigma$}_{\rm q},
\label{eq:mu}
\end{equation}
leading to the standard expressions
\begin{equation}
\mu_{\rm p}=\frac{1}{3}(4\mu_{\rm u}-\mu_{\rm d}),\,\ldots,\quad \mu_{\rm q}=
\frac{e_{\rm q}}{2m_{\rm q}}\,. \label{eq:mu_quark}
\end{equation}
The masses in the quark moments are clearly effective masses, and can be
treated as free parameters in attempting to fit the data. A 
least-squares fit to the measured octet moments, 
taken with equal weights, is shown in Table 1. The moments are given in  nuclear magnetons (nm). 
The pattern of the signs of the quark model moments agrees with observation,
while the root-mean-square deviation of 
theory from experiment is 0.14 nuclear
magnetons, or about 9\% of the average magnitudes of the moments.  
Agreement at this level can
be regarded as an outstanding success of the quark model, but the deviations 
also give a very sensitive test of baryon structure: there is presently
no completely successful first-principles theory of the moments. 
\bigskip
\begin{center}
Table 1: Quark model fit to the magnetic moments in the baryon octet.
All moments are given in nuclear magnetons.
\medskip

\begin{tabular}{|c|c|c|r|} 
\hline
Baryon & Experiment & Quark Model & $\Delta\mu$\\
\hline
$p$ & 2.793 & 2.728 & 0.065 \\
$n$ & -1.913 & -1.819 & -0.094 \\
$\Sigma^+$ & $2.458\pm 0.010$ & 2.639 & -0.181 \\
$\Sigma^-$ & $-1.160\pm 0.025$ & -0.999 & -0.161 \\
$|\Sigma^0\rightarrow\Lambda\gamma|$ & $1.61\pm 0.08$ & 1.575 & -0.03 \\
$\Lambda$ & $-0.613\pm 0.004$ & -0.642 & 0.029 \\ 
$\Xi^0$ & $-1.250\pm 0.014$ & -1.462 & 0.212 \\
$\Xi^-$ & $-0.651\pm 0.025$ & -0.553 & -0.098 \\
\hline
\end{tabular}
\end{center}
\medskip

\subsection{The Wilson-loop approach to the baryon moments}

\begin{center}
\begin{figure}[t]
\psfig{figure=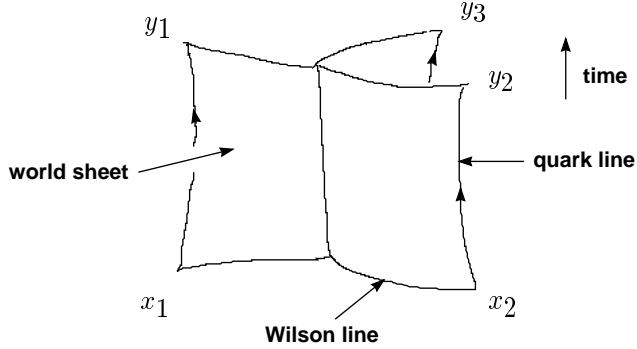,height=1.9in}
\caption{World sheet picture for the structure of a baryon. 
\label{fig:worldsheet}}
\end{figure}
\end{center}

Our approach to the baryon moment problem is based on the work of
Brambilla {\em et al.} \cite{brambilla}, who derived the interaction
potential and wave equation for the valence quarks in a baryon
from QCD using a Wilson-line construction. The basic idea is to construct a 
Green's function for the propagation  of a gauge-invariant combination of
quarks joined by path ordered Wilson-line factors
\begin{equation}
U=P\exp ig\!\int\! A_g\cdot dx, \label{wilsonline}
\end{equation}
where $A_g$ is the color gauge field. The Wilson lines sweep out a
three-sheeted world sheet of the form shown in Fig.\ \ref{fig:worldsheet}
as the quarks move from their initial to their final configurations.

By making an expansion in powers of $1/m_q$ and considering only forward 
propagation of the quarks in time using the Foldy-Wouthuysen approximation,
Brambilla {\em et al.}
are able to derive a Hamiltonian and Schr\"{o}dinger equation for the
quarks, with an interaction which involves an average over the gauge field.
That average is performed using the minimal surface approximation in which
fluctuations in the world sheet are ignored,
and the geometry is chosen to minimize the total area of the world sheet
{\em per} Wilson, subject to the motion of the quarks. The short-distance QCD
interactions are taken into account explicitly. 

The result of this construction is an
effective Hamiltonian to be used in a semirelativistic Schr\"{o}dinger
equation $H\Psi=e\Psi$,
\begin{eqnarray}
H &=& \sum_i\sqrt{p_i^2+m_i^2}+\sigma(r_1+r_2+r_3)-\sum_{i<j}\frac{2}{3}
\frac{\alpha_{\rm s}}{r_{ij}} \nonumber \\
&&-\frac{1}{2m_1^2}\frac{\sigma}{r_1}{\bf s}_1
\cdot ({\bf r}_1\times{\bf p}_1) 
+\frac{1}{3m_1^2}{\bf s}_1\cdot[({\bf r}_{12}\times{\bf p}_1)
\frac{\alpha_{\rm s}}{r_{12}^3} + ({\bf r}_{13}\times{\bf p}_1)
\frac{\alpha_{\rm s}}{r_{13}^3}] \nonumber\\
&&-\frac{2}{3}\frac{1}{m_1m_2}\frac{\alpha_{\rm s}}{r_{12}^3}
{\bf s}_1\cdot{\bf r_{12}}\times{\bf p}_2
-\frac{2}{3}\frac{1}{m_1m_3}\frac{\alpha_{\rm s}}{r_{13}^3}
{\bf s}_1\cdot{\bf r_{13}}\times{\bf p}_3+\cdots.
\label{hamiltonian}
\end{eqnarray}
Here ${\bf r}_{ij}={\bf x}_i-{\bf x}_j$ is the
separation of quarks $i$ and $j$, $r_i$ is the distance of quark $i$ from
point at which the sum $r_1+r_2+r_3$ is minimized, and ${\bf p}_i$ and
${\bf s}_i$ are the momentum and spin operators for quark $i$.
The parameter $\sigma$
is a ``string tension'' which specifies the strength of the long range
confining interaction, and $\alpha_{\rm s}$ is the strong coupling.
The terms hidden in the ellipsis include tensor and spin-spin
interactions which will not play a role in the
analysis of corrections to the moment operator, and the terms that
result from permutations of the particle
labels. The full Hamiltonian is given in \cite{brambilla}. 
This Hamiltonian, including the terms omitted here, 
gives a good description of the
baryon spectrum as shown by Carlson, Kogut, and Pandharipande \cite{kogut}
and by Capstick and Isgur \cite{isgur}, who proposed it on the basis of
reasonable physical arguments, but did not give formal derivations from
QCD.

The presence of the quark momenta ${\bf p}_i$ in the Thomas-type spin-dependent
interactions in Eq.\ (\ref{hamiltonian}) suggests that new contributions to 
the magnetic moment operator could arise in a complete theory through
the minimal substitution
\begin{equation}
{\bf p}_i\rightarrow {\bf p}_i-e_i{\bf A}_{\rm em}(x_i), \label{minimalsub}
\end{equation}
with ${\bf A}_{\rm em}(x_i)$ the electromagnetic 
vector potential associated with an
external magnetic field. However, this point is obscured in the work
of Brambilla {\em at al.} by the transformations they make to express the
equations for the Green's function in Wilson-loop form. We have therefore
repeated their derivation of the valence-quark Hamiltonian in Eq.\ (
\ref{hamiltonian}), replacing the SU(3)$_c$ color gauge field $gA_g$ by the 
full SU(3)$_c\times$U(1)$_{\em em}$ gauge field $gA_g+e_qA_{\rm em}$ 
at the beginning.
By then reorganizing the calculation of the three-quark Hamiltonian
to keep $e_qA_{\rm em}$ explicit 
throughout, and then expanding to first order in $A_{\rm em}$ with 
\begin{equation}
{\bf A}_{\rm em}=\frac{1}{2}{\bf B}\times{\bf x}_q, 
\quad {\bf B}={\rm constant,} \label{magneticpotential}
\end{equation}
we can identify the modified magnetic moment operator through the relation
\begin{equation}
\Delta H=-\mbox{\boldmath$\mu$}\cdot{\bf B}. \label{deltaH}
\end{equation}

The new moment operator, $\mbox{\boldmath$\mu$}=\mbox{\boldmath$\mu$}_{\rm
QM}+\Delta\mbox{\boldmath$\mu$}$, involves the leading corrections to the 
quark-model operator associated with the binding interactions.
$\Delta\mbox{\boldmath$\mu$}$ {\em can}, in fact, be read off from the
terms in Eq.\ (\ref{hamiltonian}) which depend on both the quark spins 
and momenta by making the minimal substitution in Eq.\ (\ref{minimalsub}).
For example, the term which depends on ${\bf s}_1\cdot{\bf r}_{12}\times
{\bf p}_1$ gives an extra contribution
\begin{equation}
\frac{e_1}{6m_1^2}{\bf x}_1\times({\bf s}_1\times{\bf r}_{12})\frac{\alpha_
{\rm s}}{r_{12}^3} \label{deltamu1}
\end{equation}
to $\mbox{\boldmath$\mu$}_1$. There are also possible orbital contributions
to the moments because the Hamiltonian mixes states with nonzero orbital
angular momenta with the ground state. These have the standard form to the
accuracy we need.

The baryon moments are now given by expressions of the form
\begin{equation}
\mu=\sum_{i=1}^3\mu_i\langle\sigma_{i,z}\rangle(1+\delta_i)+\sum_{i=1}^3
\mu_i\langle L_{i,z}\rangle, \label{mufinal}
\end{equation}
where we have quantized along $\bf B$, taken along the $z$ axis. The
expectation values are to be calculated in the baryon ground states. The
correction terms $\delta_i$ from the new operators are given for the $L=0$
baryons other than the $\Lambda$ by
\begin{eqnarray}
\delta_i &=& \frac{3\epsilon_0+\epsilon_1}{2m_1}+\frac{e_3}{e_1}
\frac{\epsilon_2}{m_3}-\frac{\Delta_0+\Delta_1}{2m_1},\quad i=1,2,\nonumber\\
\delta_3 &=& \frac{\epsilon_2}{m_3}+\frac{2e_1}{e_3}\frac{\epsilon_1}{m_1}
-\frac{\Delta_2}{m_3},\label{deltas}
\end{eqnarray}
where the $\epsilon$'s and $\Delta$'s are ground state radial matrix elements,
\begin{eqnarray}
\epsilon_0=\langle\frac{2}{3}\frac{\alpha_{\rm s}}{6r_{12}}\rangle,\quad  
& \displaystyle \epsilon_1=\langle\frac{2}{3}\frac{\alpha_{\rm s}
{\bf r}_{23}\cdot{\bf z}_2}{3r_{23}^3}\rangle,\quad 
& \epsilon_2= \langle\frac{2}{3}\frac{\alpha_{\rm s}{\bf r}_{31}\cdot
{\bf z}_3}{3r_{31}^3}\rangle \nonumber\\
\Delta_0=\langle\frac{\sigma r_{12}}{12}\rangle,\quad & \displaystyle
\Delta_1=\langle\frac{\sigma {\bf r}_{23}\cdot{\bf z}_2}{6r_{23}}\rangle,\quad 
& \Delta_2=\langle\frac{\sigma {\bf r}_{31}\cdot{\bf z}_3}{6r_{31}}\rangle.
\label{matrixelements}
\end{eqnarray}
The identical quarks in these baryons are labelled 1 and 2, the unlike quark, 
3. In writing these results, we have made the approximation $r_1+r_2+r_3\approx
\frac{1}{2}(r_{12}+r_{23}+r_{31})$, known to be reasonably accurate for the
ground state baryons \cite{kogut}, and used the corresponding Thomas spin
interaction. The result for the $\Lambda$ is similar.

\subsection{Tests of the model}

Rough estimates of the matrix elements above suggest that $\epsilon_i/m_l
\approx 0.05$ and that $\Delta_i/m_l\approx 0.3$ for the 
light-quark masses used in Refs.\ 2 and 3. Binding effects are therefore
potentially large, and are different for different quarks and baryons. To
obtain a quantitative test of these effects and reliable estimates of the
orbital contributions to the moments, expected to be small, we have 
performed a detailed analysis of the baryon wavefunctions using the 
Hamiltonian given in \cite{brambilla}. We use Jacobi-type internal coordinates
\mbox{\boldmath $\rho, \lambda$}, with $\mbox{\boldmath $\rho$}=
{\bf x}_1-{\bf x}_2$, and $\mbox{\boldmath $\lambda$}={\bf R}_{12}-{\bf x}_3$,
where ${\bf R}_{ij}$ is the coordinate of the center of mass of quarks $i$
and $j$. The most general $j=\frac{1}{2}^+$ baryon wave function for 
quarks 1 and 2 identical and orbital angular momenta $L_\rho, L_\lambda, L 
\leq 2$ has the form 
\begin{eqnarray}
\psi_{\frac{1}{2},m}&=&\left[\right.a_0\psi_a{\bf 1}
+ib_0\psi_b\,(\mbox{\boldmath$\sigma$}_1-\mbox{\boldmath$\sigma$}_2)
\cdot\mbox{\boldmath$\rho$}\times\mbox{\boldmath$\lambda$}
+c_0\psi_c\,t_{12}(\mbox{\boldmath$\rho$})\nonumber\\
&&+d_0\psi_d\,t_{12}(\mbox{\boldmath$\lambda$})
+e_0\psi_d\,(\mbox{\boldmath$\sigma$}_1-\mbox{\boldmath$\sigma$}_2)\cdot
\mbox{\boldmath$\sigma$}_3\,\mbox{\boldmath$\rho$}\cdot 
\mbox{\boldmath$\lambda$}\left.\right]\chi_{\raisebox{-.5ex}
{$\scriptstyle\frac{1}{2},m$}}^{S_{12}=1}. \label{wavefunc}
\end{eqnarray}
Here $t_{12}$ is the usual tensor operator
\begin{equation}
t_{12}({\bf x})=3\mbox{\boldmath$\sigma$}_1\cdot{\bf x}\mbox{\boldmath$\sigma$}
_2\cdot{\bf x}-\mbox{\boldmath$\sigma$}_1\cdot\mbox{\boldmath$\sigma$}_2
{\bf x}^2,
\end{equation}
and $\chi_{\raisebox{-.5ex}
{$\scriptstyle\frac{1}{2},m$}}^{S_{12}=1}$ is the standard three-particle
spinor for $S_{12}=1,\,j=\frac{1}{2},\, j_3=m$. The scalar functions $\psi_i=\psi_i(\rho^2,\lambda^2)$ are normalized together with
the accompanying spin operators. The constant coefficients $a_0,\ldots,e_0$, 
normalized to unity, give the fractions of the various component
states in $\psi$. With this form of the wave function,
we can use trace methods for the spins to calculate such quantities as
$\langle H\rangle$ and $\langle\mbox{\boldmath$\mu$}\rangle$. 

We have made variational calculations of the energies and approximate
wave functions of the ground state baryons and their first excited states
using the Hamiltonian in \cite{brambilla}. The information on the excited 
states 
allows us to calculate the coefficients $b_0,\ldots, e_0$ perturbatively.
While the Hamiltonian does mix orbitally excited states into the
$L_\rho=L_\lambda=L=0$ quark-model ground state,
the coefficients are very small, ranging from essentially zero to about
0.02 depending on the baryon. Because these coefficients only 
appear quadratically in the baryon moments, the orbital contributions to the 
moments are completely negligible.

The radial matrix elements $\epsilon_i$ and $\Delta_i$ defined in Eq.\ 
(\ref{matrixelements}) are significant, with the $\epsilon$'s ranging 
from 10 to 20 MeV, and the $\Delta$'s from 40 to 70 MeV. A
new fit to the moment data using the expressions in Eqs.\ (\ref{mufinal})
and (\ref{deltas}), with the quark masses allowed to vary, gives a small
improvement in the fitted moments, with the root-mean-square deviation
from the measured moments decreasing from 0.14\,nm for the quark model to
0.10\,nm. A secondary effect of the corrections is to change the fitted
quark masses or moments significantly, with the effective quark masses
decreasing, or the moments increasing.

We have concluded from this exercise first, that the finer details of baryon
structure are not yet described correctly in the present QCD-based model, 
and second, that the baryon magnetic moments provide a very sensitive 
check on the theory.

\section{Possible improvements in the theory}

We begin the discussion of possible improvements in the theory of the
moments by recalling the 
approximations used in the construction given by Brambilla
{\em et al.} \cite {brambilla} and in our derivation of the
moment operators:
\newcounter{approx}
\begin{list}{(\roman{approx})}{\usecounter{approx}}
\item
the $1/m_q$ expansion;
\item
the minimal surface approximation;
\item
forward propagation of the quarks in time;
\item
and the quenched approximation.
\end{list}
Of these approximations, only (iii) and (iv) are likely to affect the
moments significantly.

The $1/m_q$ expansion is to be interpreted as an expansion in 
constituent quark masses. It can be resummed in the kinetic terms in the
Hamiltonian as in Eq.\ (\ref{hamiltonian}), 
and leads elsewhere to the appearance of effective
inverse masses $1/m_i$ which need not be the same as the kinematic masses,
but represent averages of quantities such as $1/E_i=1/\sqrt{p_i^2+m_i^2}$.
Since there is no new spin dependence involved, and
we have treated the quark masses as free parameters in fitting the
moments, relativistic corrections of this sort are unlikely to change
our results.

The average over the color gauge field $A_g$ in the 
minimal surface approximation
neglects fluctuations, and minimizes the surface energy of the
world sheet to obtain the approximate Hamiltonian. Since the color fields do 
not carry charge, they do not contribute internal currents in the baryon, 
and an improved
treatment of the averaging would presumably not change the moment operator
directly. While it could change the functional form of the Hamiltonian 
somewhat, the quantitative success of the model in describing
baryon spectra suggests that the changes would not be large, and their effect 
on the moments through the $\epsilon$ and $\Delta$ parameters would be minimal.

In contrast, the remaining two approximations affect the moment operator
directly. Internal quark loops can contribute circulating currents in the
baryon. These are omitted in the quenched approximation. In addition, the
forward propagation of the quarks in time inherent in the use of the 
Foldy-Wouthuysen approximation eliminates quark pair effects connected 
with the valence lines in Fig. 1. We believe that this general quenched
picture underlies the difficulties with the model, and that it will be 
necessary to include pair effects to obtain a precise dynamical description of 
the moments. 
 
\begin{figure}[h]
\psfig{figure=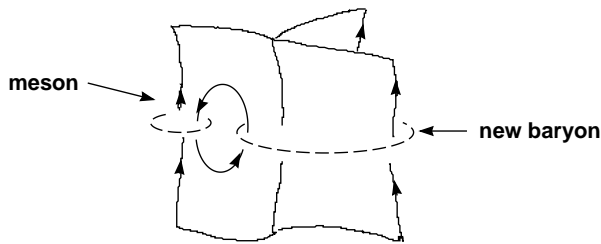,height=1.3in}
\caption{Quark loops embedded in the world sheet give meson-baryon states.
\label{fig:mesonloop}}
\end{figure}

One effect of internal quark loops can be seen in Fig.\ \ref{fig:mesonloop}.
In this figure, we suppose that a quark loop is embedded in the minimal
world sheet of the baryon. The effect is to give a meson state and a new
baryon in a world-sheet picture of meson emission and absorption.  
\noindent
An external magnetic field interacting with the system will see a meson
current as well as the baryon current, and the moment will be modified.
Meson currents were invoked in the past in attempts to explain the {\em full}\/
anomalous moments of the nucleons. Here we are only concerned with
presumably small corrections to the quark-model moments. 

\begin{figure}[h]
\psfig{figure=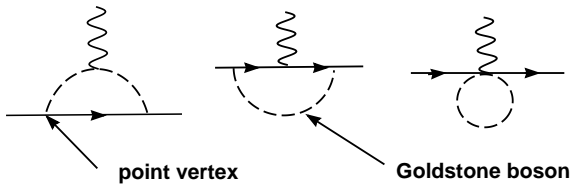,height=1.2in}
\caption{Goldstone-boson diagrams which contribution to the magnetic moments.
\label{fig:goldstone}}
\end{figure}

One approach to the calculation of meson loop effects is through
chiral perturbation theory. This has a long history, and has
been studied recently in the context of baryon moments by a number of 
authors \cite{jenkins,luty,bos}. The relevant diagrams 
for the interaction of the baryons with the pseudo Goldstone bosons
of the chiral theory are shown
in Fig.\ \ref{fig:goldstone}.
The work of Jenkins {\em et al.} \cite{jenkins}
uses heavy baryon chiral perturbation theory \cite{manohar}, and studies the
effect on the moments of the ``nonanalytic terms'' in the symmetry breaking
parameter $m_{\rm s}$. The analytic
terms in $m_{\rm s}$ are ambiguous because the inevitable
appearance of new couplings at each order in the chiral expansion,
and are ignored. The results of Jenkins {\em et al.}
are not encouraging as they stand, but we have found errors in some
of the coupling factors given in the published paper. Luty 
{\em et al.} \cite{luty}
concentrate on a simultaneous expansion in $1/N_{\rm c}$ and $m_{\rm s}$, 
and obtain interesting sum rules for the
moments but little dynamical information. Finally, Bos {\em et al.} \cite{bos}
consider the moments from the point of view of flavor
SU(3) breaking in the baryon octet
using a different chiral counting than that usually used, and obtain
a very successful parametrization for the moments. However, this model
is again nondynamical, and has seven parameters to 
describe eight measured moments. 
In fact, a fundamental problem with the chiral expansion from our perspective is that it simply parametrizes the moments with an
expansion consistent with QCD, but does not control the higher order terms
in what appears to be a slowly convergent series.

\begin{figure}
\psfig{figure=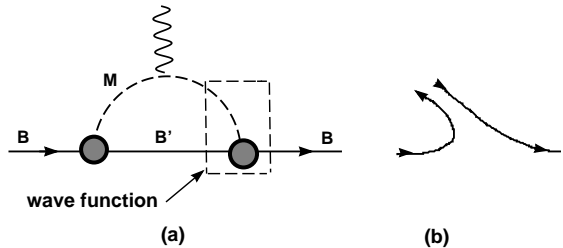,height=1.4in}
\caption{(a) Appearance of Bethe-Salpeter wave functions in the diagrams.
(b) Schematic reminder of the distributed nature of the meson-baryon vertex.
\label{fig:wavefunctionvertex}}
\end{figure}

We are presently investigating meson loop effects using a somewhat 
different approach suggested by the world-sheet picture. The baryon appears in 
this picture as an extended object which must absorb the recoil momentum
in the emission of a meson. We would therefore expect wave function effects
(form factors) to be important for high meson momenta, and to supply a natural
cutoff for loop graphs. The wave function appears naturally when the 
process is viewed using ``old fashioned'' instead of Feynman perturbation
theory, as indicated in Fig.\ \ref{fig:wavefunctionvertex}: energy
denominators and vertex functions combine to give the $B'M$ component of
the wave function for a baryon $B$, or, with different time ordering, the 
$BM$ component of the wave function of $B'$. Since the wave functions are
expected to be fairly soft, with characteristic momenta below the
chiral cutoff of $\sim 1$ GeV, they can be expected to combine a number of
terms in the chiral expansion in a way that is dynamically accessible
through approximate models derived from QCD such as that of Brambilla 
{\em et al.} \cite{brambilla}. The non-point character of the meson-baryon vertex in spacetime indicated in Fig.\ \ref{fig:wavefunctionvertex} 
contrasts sharply with the point vertex used in chiral
perturbation theory, Fig.\ \ref{fig:goldstone}, and is also likely to play
a role. We have reached the same conclusions by 
studying the exact sideways dispersion relations for the moments given by
Bincer \cite{bincer}. The problem there is in extracting the quark-model
moments. 

We note finally that a different aspect of symmetry breaking, the 
suppression of strange-quark loops through mass effects, appears as a
natural possibility in a world-sheet picture. Note that this is {\em not}
the same as the suppression of kaon loops considered by other authors; see,
e.g., \cite{jenkins} and the references given there.

\section{Conclusions}

On the basis of the work sketched above, we have concluded that the magnetic 
moments of the baryons give a sensitive test of baryon structure and of
approximations in QCD. In particular, the quenched approximation, while reasonably successful when used in the calculation of
baryon and meson spectra, appears to fail for the moments. The accurate
calculation of the moments by lattice methods will presumably involve
going beyond that approximation, and will provide a precision test of
the methods used.

We find also that the world-sheet picture of baryon structure gives useful
insights into the moments problem, and provides a new point of view which
could be developed further. Problems which need further study within this 
approach to QCD include the following: 
\begin{list}{(\roman{approx})}{\usecounter{approx}}
\item
establishing the connection to, and the relevance of, the chiral limit;
\item
the development of methods to incorporate loop effects which build in
the extended structure of the baryons; 
\item
and the possible usefulness of string theory methods in the calculation
of loop effects.
\end{list}

\section*{Acknowledgments}
The authors have benefited from conversations with Dr.\ Nora Brambilla,
and appreciate her interest in this work, and her organization
of the Como Conference. 
This research was supported in part by the U.S. Department of Energy under
Grant No.\ DE-FG02-95ER40896, and in part by the University of Wisconsin
Graduate School with funds from the Wisconsin Alumni Research Foundation.

\end{document}